# ODT FLOW: A Scalable Platform for Extracting, Analyzing, and Sharing Multi-source Multi-scale Human Mobility

Zhenlong Li[1*], Xiao Huang[2], Tao Hu[3], Huan Ning[1], Xinyue Ye[4], Xiaoming Li[5]

[1]Geoinformation and Big Data Research Lab, Department of Geography, University of South Carolina, Columbia, SC, USA

[2]Department of Geosciences, University of Arkansas, Fayetteville, AR, USA

[3]Center for Geographic Analysis, Harvard University, Cambridge, MA, USA

[4]Department of Landscape Architecture & Urban Planning, Texas A&M University, TX, USA

[5]Department of Health Promotion, Education, and Behavior, University of South Carolina, Columbia, SC, USA

[*]Email: zhenlong@sc.edu

**Abstract**

In response to the soaring needs of human mobility data, especially during disaster events such as the COVID-19 pandemic, and the associated big data challenges, we develop a scalable online platform for extracting, analyzing, and sharing multi-source multi-scale human mobility flows. Within the platform, an origin-destination-time (ODT) data model is proposed to work with scalable query engines to handle heterogenous mobility data in large volumes with extensive spatial coverage, which allows for efficient extraction, query, and aggregation of billion-level origin-destination (OD) flows in parallel at the server-side. An interactive spatial web portal, ODT Flow Explorer, is developed to allow users to explore multi-source mobility datasets with user-defined spatiotemporal scales. To promote reproducibility and replicability, we further develop ODT Flow REST APIs that provide researchers with the flexibility to access the data programmatically via workflows, codes, and programs. Demonstrations are provided to illustrate the potential of the APIs integrating with scientific workflows and with the Jupyter Notebook environment. We believe the platform coupled with the derived multi-scale mobility data can assist human mobility monitoring and analysis during disaster events such as the ongoing COVID-19 pandemic and benefit both scientific communities and the general public in understanding human mobility dynamics.

**1. Introduction**

Reliable monitoring of human mobility, referring to the movement of human beings (individuals as well as groups) in space and time, plays a fundamental role in a variety of fields, such as tourist management (Lamsfus et al., 2015; Hall, 2005), migration (Sirkeci and Cohen, 2016; Afifi et al., 2016), urban planning (Hillier et al., 2009; Bhat et al., 2004; Wu et al., 2014), demand forecasting (Kitamura et al., 2020; Smolak et al., 2020), disaster management (Martin et al., 2020; Jiang et al., 2019), and epidemic modeling (Kumar et al., 2020; Olive et al., 2020; Ye, Li, and Peng, 2020), to name a few. The ongoing COVID-19 pandemic (as of the time of writing) uniquely highlights the necessity of human mobility monitoring in a rapid and comprehensive manner.

Traditionally, human mobility studies rely on either aggregated and temporally sparse governmental statistics (e.g., census data) (Santos et al., 2011) or selective, small-scale surveys (e.g., local travel survey and evacuation survey) (Barbosa et al., 2018; Olabarria et al., 2013). While these well-documented mobility records from official statistics and surveys facilitate our understanding of spatial interactions, their intrinsic limitations stand out: the former provides only coarse-grained spatiotemporal human movement patterns while the latter is limited by the spatial scale. With the advent of the Global Positioning System (GPS) technologies, game-changing data sources of fine-scale human mobility become available. Fostered by the concept of "Citizen as Sensors" (Goodchild, 2007) and the popularity of mobile devices, timely geospatial information can be extracted from the enormous sensing network constituted by millions and even billions of mobile devices holders in both passive (e.g., wireless networks and cellphone GPS) and active (e.g., navigation services and social media posts) fashions. Despite the availability of mobility data on a fine-grained spatiotemporal scale, new challenges start to appear, as such detailed human mobility records fall in the category of Big Data that can be characterized by the 5Vs, i.e., *Volume*, *Velocity*, *Variety*, *Veracity*, and *Value* (Yang et al., 2017; Wang et al., 2016; Marr, 2015; Li, 2020), thus demanding a paradigm shift of data handling approach from a traditional static paradigm to an accelerating arena.

Digitally traced human mobility records are generally with a large *Volume*. For example, the size of daily geolocation reports from smartphones and other mobile devices are in GB- and even TB-level (Warren and Skillman, 2020). In addition, massive social media users produce hundreds of millions of posts every day (Li et al., 2018; Yang et al., 2017; Ye et al., 2020). Although only a small proportion of them are geotagged, such an amount already exceeds the capability of traditional data handing approaches and demands improved strategies to tackle the computing, storing, and analyzing issues. *Velocity* refers to the fast generation of human mobility data, as human mobility records from digital devices are constantly being produced. Such a high velocity in the data flow demands the capability of cyberinfrastructures to organize, summarize, visualize, and analyze data in a rapid manner (Khan et al., 2014). *Variety*, on the one hand, points to the multi-faceted nature of human mobility originated from its sources whose population penetration rates and covered population spectrums are not necessarily the same. Studies revealed that mobility records from different sources present a certain level of similarity in general, however, with notably unique and even contrasting characteristics, reflecting human mobility from varying yet valuable perspectives (Huang et al., 2021a; Hu et al., 2021). On the other hand, *Variety* also highlights the multi-scale nature of mobility data. During the COVID-19 pandemic, for example, mobility records are reported at a variety of levels that include the country (Huang et al., 2020a; da Câmara Ribeiro-Dantas et al., 2020), state/province (Descartes Labs Aggregated Mobility Index, 2020), county (Apple Mobility Trends Reports, 2020; Google Community Mobility Reports, 2020), and even finer census levels (SafeGraph Social Distancing Metrics, 2020). *Veracity* refers to the biases, noises, and abnormality (Rubin and Lukoianova, 2013), to list a few, which can be traced back to the source provenance in mobility datasets. In the context of human mobility datasets, *Veracity* not only demands proper cleaning, preprocessing, and aggregating procedures, but also highlights the importance and necessity of fusing and sharing multiple data sources for human mobility studies. Last but not least, *Value* refers to the insights derived from human mobility patterns to support decision-making and policy implementation. As the knowledge of human mobility patterns plays an essential role in a wide range of research domains, designing platforms that facilitate human mobility data analysis and sharing is much needed, especially during disaster events like the COVID-19 pandemic that requires a rapid sharing of such knowledge. In tandem with the challenges of 5*V*s in big mobility data is the growing awareness of reproducibility and replicability (Choi et al., 2021; Sui and Kedron, 2020), as the extensive usage of locational data in various domains uniquely highlights such concerns. The 5*V* challenges, coupled with the necessity of facilitating reproducibility and replicability, largely raise the bar for not just mobility visualization platforms but data sharing platforms in general.

The COVID-19 pandemic has fostered many open-sourced mobility datasets and online mobility visualization platforms, as scholars realize the importance of fast insights from mobility patterns for better, timely decision-making. Some notable open-sourced mobility datasets include social media derived mobility datasets, Apple/Google mobility reports, SafeGraph social distancing metrics, Cuebiq Mobility Index, Baidu Mobility Index, Descartes Labs Mobility Index, University of Maryland Mobility Metrics and Social Distancing Index, and Unacast Social Distancing Index. These datasets differ in sources, spatiotemporal scales, geographic coverages (Hu et al., 2021). In terms of online visualization platforms, one notable effort is by Gao et al. (2020), who designed a dashboard to present mobility dynamics at the U.S. county level using mobility records from Descartes Labs. Cuebiq developed a mobility dashboard featured by the visualization of Cuebiq's Mobility Index (https://www.cuebiq.com/visitation-insights-covid19/), which enables brands, researchers, government agencies, and health professionals to understand shifts in mobility trends across the U.S. at multiple levels (nationally, state-wide, county and industry vertical). In collaborating with COVID-19 Mobility Data Network (https://www.covid19mobility.org/), Stamen, a data visualization and cartography design studio, established a visualization platform that provides mobility insights at the country level using aggregated population movement data from Facebook's Data for Good program. However, report-based mobility data (e.g., Google and Apple mobility reports), despite that they can be easily handled, do not provide origin-destination (OD) flows that are essential for certain location-based studies. In addition, dashboards that facilitate OD visualization often lack flexibility, as most dashboards are constructed from pre-computed statistics with highly aggregated results, posing challenges for their practicality in multi-scale studies. After reviewing existing efforts in open-source mobility datasets and visualization platforms, we notice that few frameworks have been designed with the capacity to integrate/compare multi-source and multi-scale global mobility data as well as to provide interactive query, visualization, downloadable, and programmatic access options according to user-defined spatiotemporal restrictions (e.g., Al-Dohuki et al., 2019).

In response to the soaring needs of human mobility data during the COVID-19 pandemic and to fill the aforementioned gaps, we develop a scalable online platform for extracting, analyzing, and sharing multi-source multi-scale human mobility data. At the core of the platform is an origin-destination-time (ODT) data model designed to work with scalable query engines to handle heterogenous mobility data in large volumes with extensive spatial coverage, which allows for efficient extraction, query, and aggregation of billion-level OD flows in parallel at the server-side. Built upon the ODT model and scalable query engines, an online ODT Flow Explorer is developed that allows researchers to explore multi-source mobility datasets with flexible configurations. In response to the challenges in reproducibility and replicability, we develop ODT Flow REST APIs that provide researchers with the flexibility to access the data programmatically via workflows, codes, and programs. In addition, we provide case studies to demonstrate the usage and potential of the APIs along with scientific workflows and Jupyter Notebooks to facilitate on-demand mobility data access, analysis and visualization.

## 2. Methodology

### 2.1 System design

The ODT Flow platform consists of five layers, including the data source layer, data processing and management layer, web server layer, user interface layer, and user community layer. The system is designed to tackle the big data challenges in a scalable and open computing environment, aiming to serve as a bridge to connect heterogeneous big human movement data sources (the first layer) to user communities for research and applications (the fifth layer). The movement data sources include data that contain or can be used to extract OD flows with a time interval such as daily. Such data sources include, for example, geotagged Twitter data, mobile phone data, and transportation data. Different from the index-based mobility data that are highly aggregated without OD information, these data sources are big in volume, variety or heterogeneous in format and spatiotemporal resolutions, and noisy and inconsistency in data quality, which highlights the well-defined Big Data challenges (Hu et al., 2021). Besides, these data also pose shareability, reproducibility, and replicability challenges in the Big Data era that draw attention in the scientific community (Kedron et al., 2021).

In the system, the *volume* challenge of the data is addressed in a scalable distributed database (HDFS, Hadoop Distributed File System), coupled with a spatiotemporal partition mechanism for efficient data query and access. The *variety* challenge is handled with an ODT data model by extracting spatiotemporal standardized OD flows from heterogeneous movement data sources to construct a unified ODT cube. The computing- and data-intensive OD flow extraction process is carried out in a parallel computing cluster using the data source-specific human mobility extraction algorithms. The ability to integrate multi-source mobility data also helps tackle the *veracity* challenge, as fusing multiple mobility sources can provide a more comprehensive picture of human mobility (Huang et al., 2021a). The shareability and usability of the big movement data are facilitated with a spatial web portal allowing for interactive query, visualization, and download of ODT flows. RESTful Application Programming Interfaces (ODT Flow REST APIs) are developed to allow users to query and access the data in a flexible programmatical way using scientific workflow and Jupyter notebook, to list a few. Accessing data with APIs can help to achieve research reproducibility and replicability (Choi et al., 2021).

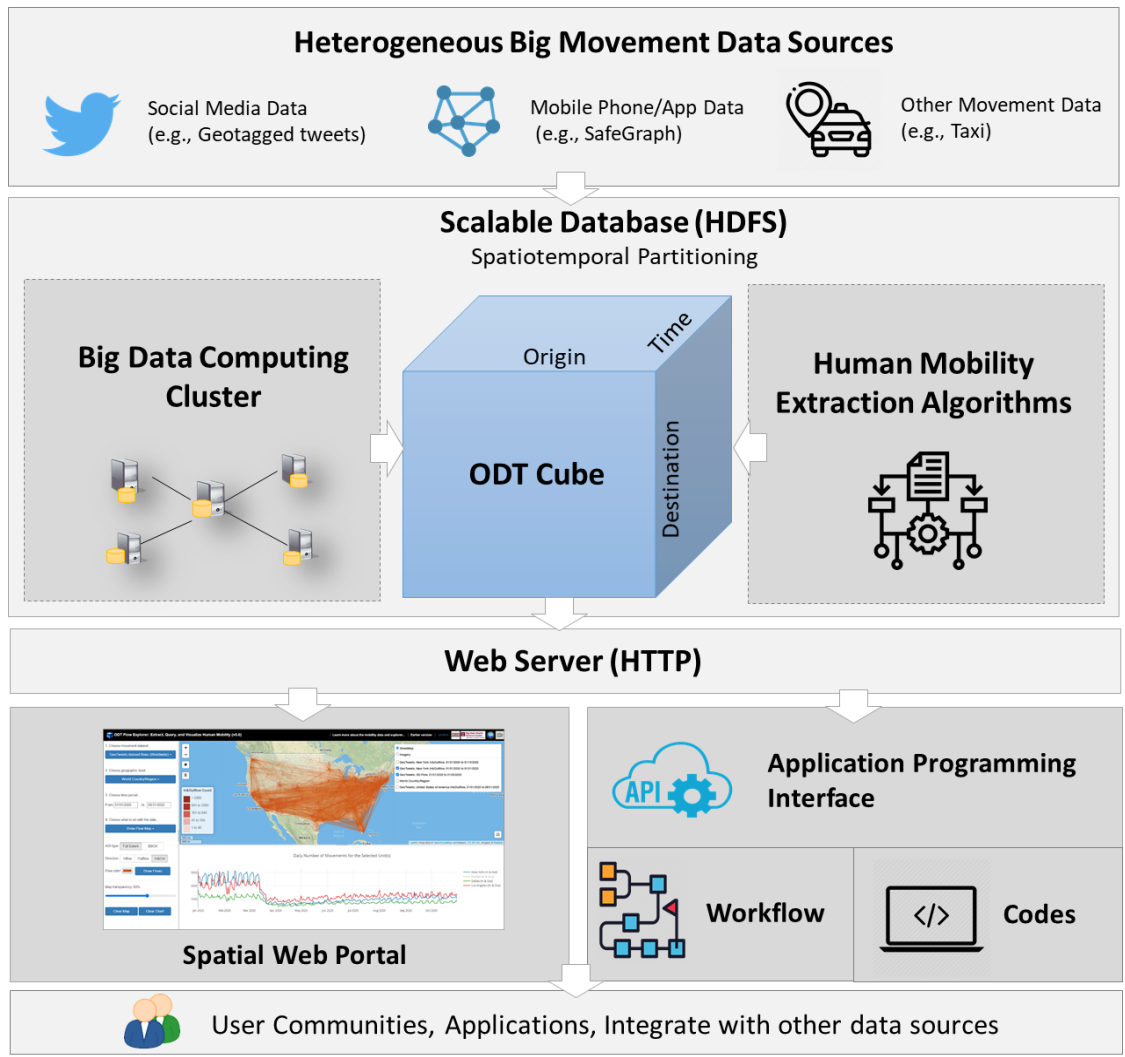

Figure 1. The system architecture of the ODT Flow platform.

**2.2 ODT cube**

The ODT cube, a conceptual data model, is designed to work with a scalable computing cluster to efficiently manage, query, and aggregate billions of OD flows at different spatial and temporal scales. The ODT cube was previously introduced in our research proposal (Li et al., 2020) and is implemented in this study. Data cube has been widely used to model and visualize multi-dimension and multi-scale data (Stolte et al., 2003; Guo et al., 2006; Li et al., 2016). In the geospatial domain, a special data cube called Space-Time Cube is often used to model time series spatial data and visualize movement data (Kraak, 2003; Kveladze et al., 2015; Yang et al., 2017). Different from the space-time cube, where space is treated as two dimensions (X and Y, or latitude and longitude) in the cube, the ODT cube is place-based with a time dimension added to the OD matrix. ODT cube does not directly capture the movement trajectory, as the OD matrix for each timestamp is independent, recording the number of population flows between places during a specific time period (e.g., in an hour or day).

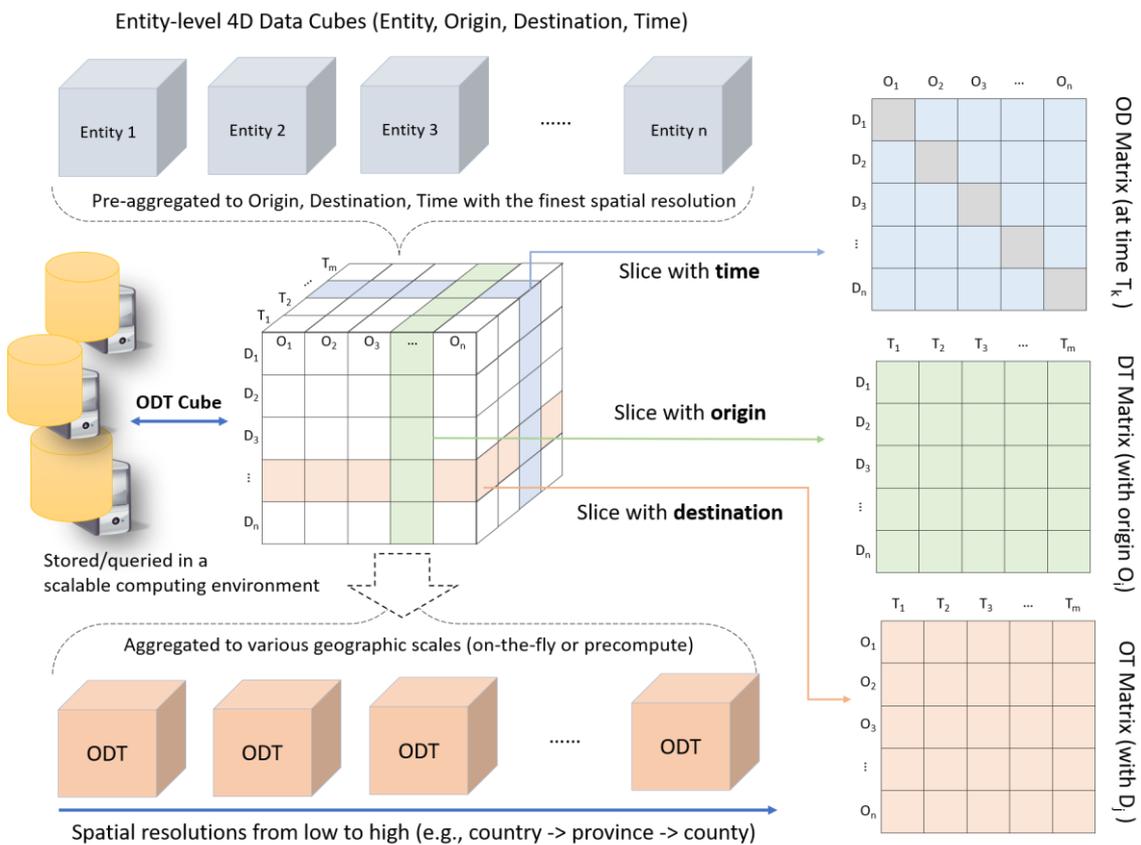

Figure 2. Illustration of Origin-Destination-Time Cube for big OD data query analytics.

To construct the ODT cube, an individual level 4D data cube (entity, origin, destination, time) is first constructed for each entity to extract the individual OD flows from millions or billions of data points. The entity can be a social media user, a cell phone user, or a place if individual flows are not available. As different movement data sources have different data formats, structures, and spatiotemporal resolutions, a data source-specific algorithm/program is needed to extract individual OD flows. Section 2.3 elaborates our approaches of extracting OD flows from Twitter data and SafeGraph data. The 4D data cubes are then aggregated along the origin, destination, and time dimensions to construct the ODT cube. Each cell of the ODT cube records the number of people that moved from the origin location to the destination location during a time period. The ODT cube can be further used to derive cubes at different geographic scales using spatial aggregation. This process can be performed on-the-fly based on the request or pre-computed and cached to optimize the performance. By slicing the ODT cube from different dimensions, three types of matrices can be derived: OD matrix recording population flows between places during a time period; destination-time (DT) matrix recording population inflows; and origin-time (OT) matrix recording population outflows. The diagonal cells of the OD matrix record the intra-movements for a place (O = D) during a time period.

Constructing the ODT cube needs to perform point-in-polygon spatial operations for billions of data points and thus is extremely data- and computation-intensive. For example, creating an ODT cube using 2019 geotagged Twitter data needs to conduct point-in-polygon checking for 1.4 billion geotagged tweets (posted by 17 million unique Twitter users) against hundreds or thousands of places depending on the geographic level. To address this challenge, the cube building process is carried out in a scalable parallel computing environment based on a stack of open-source solutions, including Apache HDFS, Hive, Impala, and GIS Tools for Hadoop by Esri. The environment can be configured and reproduced using the open-source Cloudera Distribution Hadoop (Cloudera, 2021) either on an on-premise computing cluster or cloud-based computing cluster. HDFS is used as a scalable data storage for big movement datasets. Hive, coupled with GIS Tools for Hadoop, is used for performing spatial operations in parallel for billions of data points to extract OD flows. The generated ODT cube, which may contain billion of cells, is stored in HDFS as a big table distributed across computer nodes.

**2.3 Human mobility extraction**

In this study, we generate ODT cubes by extracting population flows from two data sources: worldwide geotagged tweets collected using the Twitter public API (Twitter, 2021), and Social Distancing Metrics (SDM) provided by SafeGraph based on U.S. mobile devices (SafeGraph, 2020). As these two data sources are different in data structure, format, and spatiotemporal resolution, we develop two computational approaches for the data sources, respectively.

*2.3.1 ODT cube construction from Twitter data*

To generate the ODT cube from Twitter data, two types of OD flows are extracted and combined from geotagged tweets capturing Twitter users' single-day movement and cross-day movement. The concept of single-day and cross-day movements is introduced in Huang et al. (2020a). In general, the single-day movement represents the users' daily maximum travel distance of all locations relative to the initial location, and cross-day movement measures the mean center shift between two consecutive days. In this study, instead of computing the movement distance as Huang et al. (2020a), we are interested in each user's origin location and destination location on a daily basis to construct the entity-level 4D cube, denoted as (user, origin, destination, day). Note that the Twitter-derived OD flows do not consider users' home location. The movements were directly derived from the locations of geotagged tweets at the user level on a daily basis. Based on the 4D cube, ODT cubes are derived by aggregating the individual flows at specific geographic levels (e.g., county, state, world first-level subdivision, or country) as explained in section 2.2. Non-human tweets (posted by bots, such as weather reports and job advertisements) need to be filtered out when computing the OD flows as they are irrelevant to human mobility (Martin et al., 2020). These tweets are removed by checking the tweet source. For example, tweets automatically posted for jobs from the source TweetMyJOBS were excluded. A list of Twitter sources that indicate human-posted tweets and the descriptive statistics of the worldwide geotagged tweets collected using the Twitter public API in 2019 is detailed in Li et al. (2021, Appendix B).

*2.3.2 ODT cube construction from SafeGraph data*

To generate the ODT cube using SafeGraph data, we first extract the daily OD flows from SDM (SafeGraph, 2020). There are 23 fields in the SDM table, and we use 3 of them to extract the daily population movement, including *origin_census_block_group*, *destination_cbgs*, and *date_range_start*. The *origin_census_block_group* is the unique 12-digit FIPS (Federal Information Processing Standards) code for the Census Block Group (cbg). *destination_cbgs* contains a list of key-value pairs with the key indicating the destination census block group (from the origin census block group) and "value is the number of devices with a home in *census_block_group* that stopped in the given destination census block group for >1 minute during the time period" (https://docs.safegraph.com/docs/social-distancing-metrics). The *date_range_start* was used to extract the date information. Based on the three fields, we generate the entity-level 4D cube, denoted as (cbg, origin, destination, day), with each cell showing the number of devices from an original block group to a destination block group on a daily basis. Note that the SafeGraph-derived OD flows consider devices' home location (the movements are originated from home). For example, a flow of 100 devices (users) from county A to county B indicates that the home location of the 100 devices is in county A. Based on the 4D cube, ODT cubes are derived by aggregating the block group level flows into other geographic levels (e.g., county and state) in the U.S.

The processes of constructing the ODT cubes from Twitter data and SafeGraph data are carried out using Impala and Hive coupled with Esri GIS tools for Hadoop on the scalable computing environment. New computational approaches can be developed to create ODT cubes from other movement data sources following the two examples.

**2.4 ODT cube-based human mobility analysis**

Once the ODT cube is built, the traditional data cube atomic operations including slice, dice, drill up, drill down, and pivot can be applied to conduct various spatiotemporal query, analysis, extraction, and aggregation of billions of flows to support interactive exploration of human mobility at various geographic scales. These atomic cube operations are detailed in Datta and Thomas (1999). For example, the slice operation can be used to derive three matrices (OD, OT, and DT) from ODT cube to capture different aspects of the flows. From the performance perspective, as the ODT cube is stored and managed in HDFS, these cube operations can be efficiently conducted using Structured Query Language (SQL) with Apache Impala, an open-source parallel computing engine that offers low latency and high concurrency for analytic queries on Hadoop. Here, we provide four application scenarios to illustrate how ODT cube coupled with the cube operations can help in exploring and analyzing mobility data (Figure 3).

- Computing the number of population flows (inflow, outflow, or both) between a place (e.g., county) and all other places in the study area during a selected time period. This information can be applied in infectious disease modeling and other applications that benefit from the knowledge of population movement between places. This computation can be achieved by slicing the ODT cube to create an OT or DT matrix and then perform a temporal aggregation with the selected time period. The result can be visualized as choropleth maps.

- Computing the number of daily movements (inflow, outflow, or intraflow) for a place. This information reveals population movement trend over a time period, which can be used to examine, for example, the response to stay-at-home orders during the COVID-19 pandemic. This computation can be achieved by slicing the ODT cube to create an OT or DT matrix and then perform a spatial aggregation with the selected place. The result can be visualized as time series charts.

- Computing the number of population flows among all places (or selected places) in the study area during a specific time period. This computation can be achieved by dicing the ODT cube and aggregate along the time dimension. The result is an OD matrix that can be visualized as flow maps.

- Extracting a subset of the mobility data based on the user-defined criteria including interested area, geographic levels, time period, and so on. This operation can be achieved by dicing the ODT cube into a subcube. The result can be returned as csv (comma-separated values) data files for further analysis. The ability to extract interested subcubes on-the-fly is essential as it is often challenging and time-consuming to download the entire big movement dataset to extract the needed information.

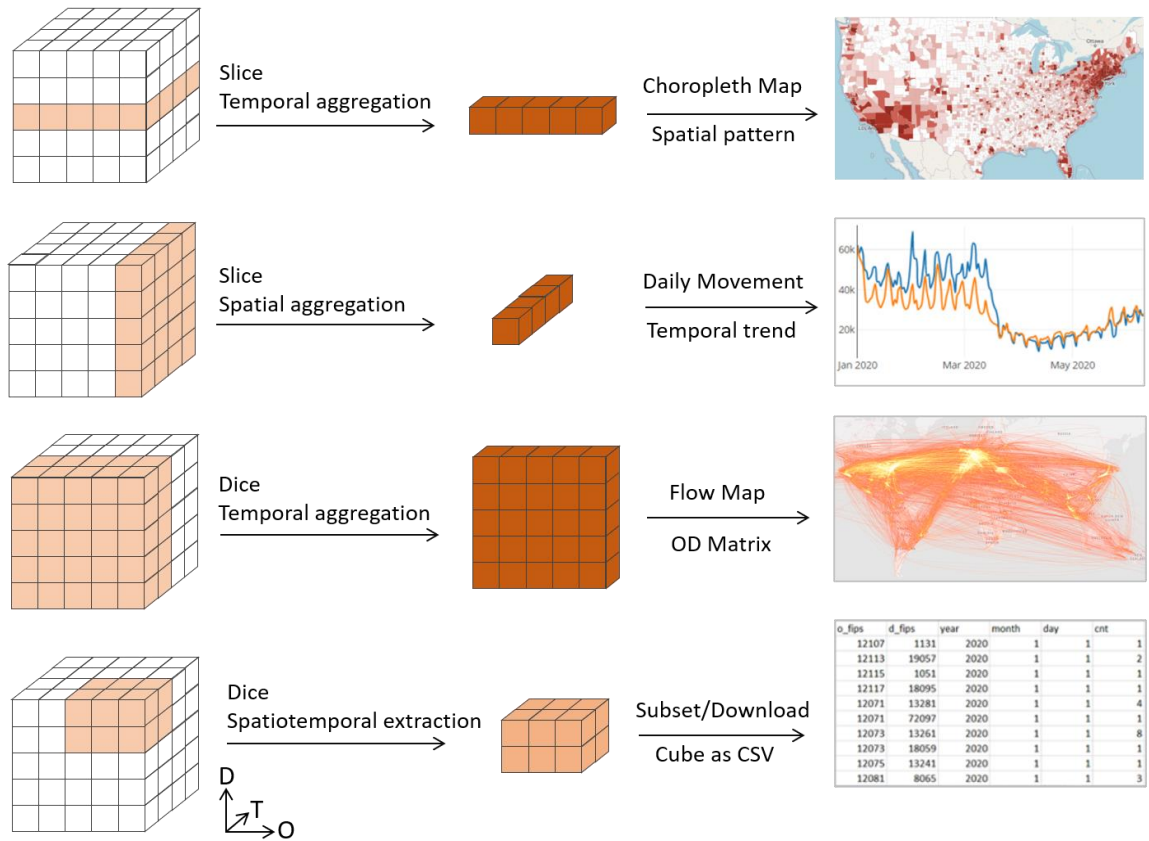

Figure 3. Four application scenarios exemplifying how ODT cube coupled with the traditional cube operations and a scalable parallel computing environment can help explore big mobility data.

## 3. Results

### 3.1 Global multi-scale daily OD flows

In this study, the daily OD flows in 2019 and 2020 were extracted using worldwide geotagged tweets collected with Twitter public API and U.S. SafeGraph social distancing metrics data downloaded from SafeGraph website (Table 1). For Twitter data, about 2.7 billion geotagged tweets (excluded non-human tweets) posted by about 25 million Twitter users were used in the flow extraction computation, resulting in about 637 million entity level daily OD flows (user, O, D, T). For SafeGraph data, over 160 million social distancing metrics records were used in the flow extraction computation, resulting in over 11 billion entity level daily OD flows (cbg, O, D, T).

The entity level OD flows were further aggregated into ODT cubes with five geographic scales, including world country/territory (Twitter), world first-level subdivision (Twitter), U.S. state (Twitter and SafeGraph), U.S. county (Twitter and SafeGraph), and U.S. census tract (SafeGraph). The number of daily OD flows (cube cells) for each geographic level and data source are listed in Table 1. These multi-scale OD flow datasets can be queried, extracted, and visualized using the ODT Flow Explorer interactively and the ODT Flow REST APIs programmatically.

Table 1. Statistics of the derived daily OD flows from Twitter data and SafeGraph data

|  | Twitter-derived OD Flow | SafeGraph-derived OD Flow |
| --- | --- | --- |
| Spatial coverage | Worldwide | U.S. |
| Temporal coverage[1] | 2019-2020 (daily) | 2019-2020 (daily) |
| Original data records | 2,695,552,594 geotagged tweets by 24,863,844 Twitter users | 160,301,510 SafeGraph social distancing metrics records |
| Derived Entity-O-D-T | 636,984,772 | 11,108,696,071 |
| World country/territory | 1,253,291 | — |
| World 1$^{st}$-level division | 9,333,761 | — |
| U.S. state | 809,741 | 1,958,450 |
| U.S. county | 10,206,119 | 439,790,381 |
| U.S. census tract | — | 6,710,889,890 |

[1]Selected time period in this study

**3.2 ODT Flow Explorer**

The ODT Flow Explorer is an interactive spatial web portal for on-demand querying, slicing, aggregating, and visualizing the multi-scale daily population flows with a few clicks (Figure 4). The front-end of the portal was developed with JavaScript, Html, and CSS using a stack of open-source libraries including jQuery (https://jquery.com), Bootstrap (for web interface, https://getbootstrap.com), and Leaflet (for mapping, https://leafletjs.com). The back-end of the portal was developed using Java and is hosted by Apache Tomcat (http://tomcat.apache.org). The portal is connected to query engines (Hive and Impala) on a scalable computing cluster to access and analyze billion-level OD flows in parallel. The web portal has attracted over 2,200 visits from 31 countries according to the RevolverMaps' real-time visitor statistics widget (RevolverMaps, 2021) as the time of writing.

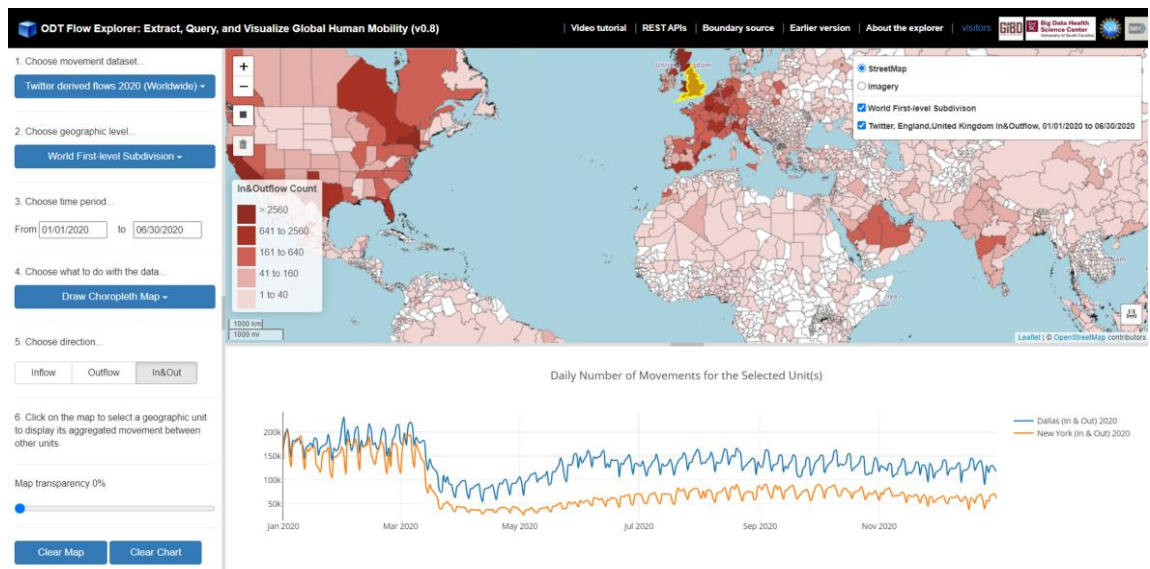

Figure 4. The user interface of the interactive spatial web portal. The world county and world first-level subdivision boundary data are derived from https://gadm.org.

**3.3 ODT Flow REST APIs**

Representational State Transfer (REST) is a widely used architectural style for developing web applications and web services (Fielding, 2000). The ODT Flow REST APIs provide functions for on-demand query, extraction, and aggregation of ODT cube, and deliver the results as a flat data table to be used by other applications in a programming way. Three types of APIs have been implemented in this study at the time of writing: 1) get the aggregated movement between a selected place and other places during a time period, 2) get the daily inter-unit movements between the selected place and other places or the selected place's daily intra-unit movements, and 3) extract the OD flows in either temporally aggregated format (OD matrix) or daily format (ODT cube) for a specified geographic area and time period.

As the APIs allow access billion-level flows programmatically, they can be integrated with other computing environments such as KNIME workflows and Jupyter Notebooks to develop sharable and reproducibility analyses that involve human mobility, which is demonstrated with the case studies in the following section. More APIs will be added in the future development of the ODT Flow platform.

## 4. Demonstration

### 4.1 Using the ODT Flow Explorer to explore, visualize, and download human mobility at different geographic scales

To use the portal, users start by choosing a mobility dataset, a geographic scale, and a time period, and then choose what to do with the selected data. Four options are available (as of version 0.8): Choropleth Map, Flow Map, Daily Movements, and Download, corresponding to the four application scenarios explained in section 2.4.

The *Choropleth Map* function visualizes a place's aggregated flows between other places during a time period as a choropleth map. Three types of flow directions can be configured: *Inflow*, *Outflow*, and *In&Out*. *Inflow* refers to the number of users/devices from other places moving to the selected place during the selected time period. *Outflow* refers to the number of users/devices moving from the selected place to other places. *In&Out* contains the movements from both directions. Figure 5a, b show the SafeGraph-derived county population flows to New York County (Manhattan) from 03/08/2020 to 03/14/2020 and for the following week (03/15/2020 to 03/21/2020). Figure 5c, d show the Twitter-derived flows between England, UK and other first-level administrative units in the Europe area for 01/01/2020 - 02/29/2020, and 03/01/2020 - 04/30/2020. The impact of the COVID-19 pandemic on human mobility can be observed in both areas and scales.

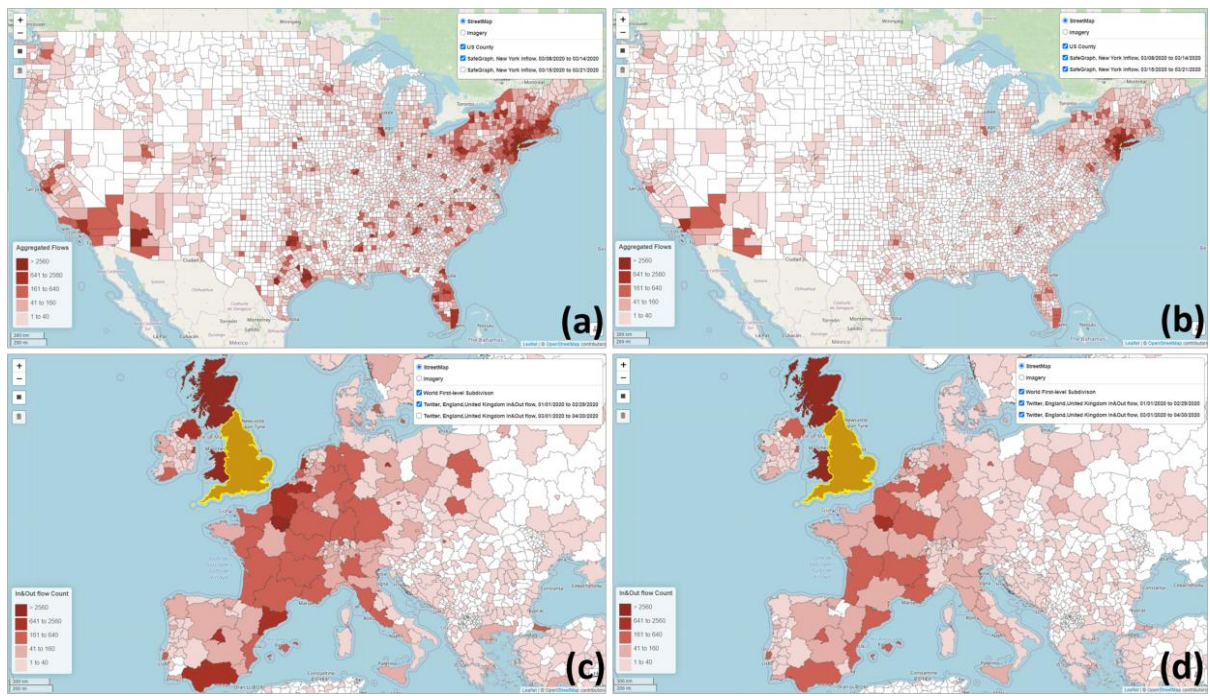

Figure 5. SafeGraph-derived county population flows to New York County from (a) 03/08/2020 to 03/14/2020 and (b) for the following week (03/15/2020 to 03/21/2020). Twitter-derived in & out flows between England, UK and other first-level

administrative units in the Europe area for (c) 01/01/2020 to 02/29/2020, and (d) 03/01/2020 to 04/30/2020

The *Flow Map* function visualizes the OD lines based on the selected dataset, geographic level, and time period. Users can choose the area of interest (AOI) by drawing a bounding box on the map or use the full spatial coverage of the data. Flow direction (*Inflow*, *Outflow*, and *In&Out*) and flow color can also be configured. The width of each flow is weighted based on the number of device/user movements for display only. Figure 6 shows county-level population movement from 01/01/2020 to 01/05/2020 derived from Twitter (a) and SafeGraph (b). Note that the flow map function aims to provide a quick overview of the selected flows. To interactively visualize massive flows, the ODT Flow REST APIs can be used along with kepler.gl (a WebGL-enabled mapping library) in the Jupyter Notebook environment (see section 4.3.2).

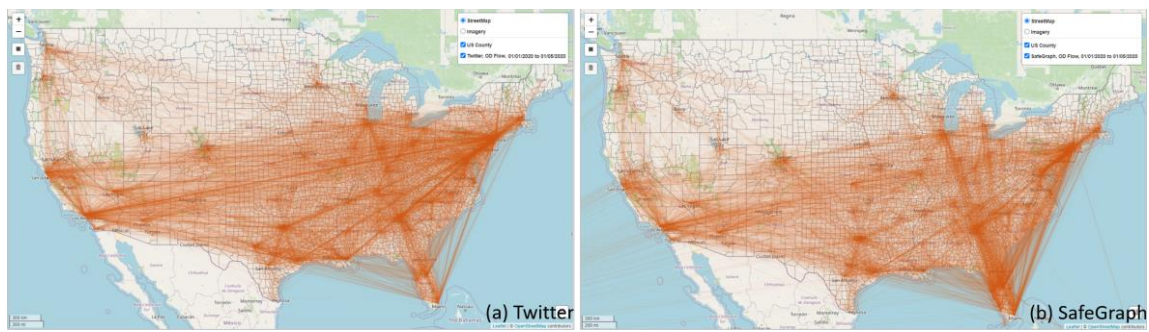

Figure 6. County-level population flows from 01/01/2020 to 01/05/2020 derived from (a) Twitter and (b) SafeGraph. Note: For SafeGraph-derived mobility data, only flows with aggregated device number great than 20 within the selected time period are displayed for performance consideration.

The *Daily Movements* function shows the trend of the daily movement for a selected place. Besides *Inflow*, *Outflow*, *In&Out*, the direction option also includes

*Intraflow,* which indicates the number of daily movements within the selected place (flows with a movement distance greater than zero but not crossing the unit boundary). Figure 7 demonstrates the daily population movement trend revealed in different geographic scales of the census tract, county, and country from 01/01/2019 to 12/31/2020. Figure 7a shows the country level daily intra-movements for Spain and Argentina (based on Twitter-derived OD data). The two countries exhibit a similar mobility reduction trend in March and April 2020 but show different recovery patterns starting from May 2020: Spain's mobility gradually rebounded back until mid-August when it started to drop again; Argentina's mobility has remained at a relatively low level. Figure 7b shows the U.S. county-level daily inflows to New York County from 01/01/2019 to 12/31/2020 (based on SafeGraph-derived OD data). In both geographic scales and locations, the impact of the COVID-19 pandemic on human mobility is well reflected by the sharp drop around March 11, 2020. Figure 7c shows the daily intraflow and interflow trend for two census tracts in Columbia, South Carolina from 01/01/2019 to 12/31/2020: one tract is part of the campus of the University of South Carolina (Figure 7c) and the other one is a residential area (Figure 7d). For the campus tract, the normal summer holiday season (May 15 – August 31) in 2019 is revealed, and the campus closures due to COVID-19 are also reflected by the extremely low mobility from March to August in 2020. The residential tract, on the other hand, shows no obvious seasonal pattern and the impact of COVID-19 is less dramatic.

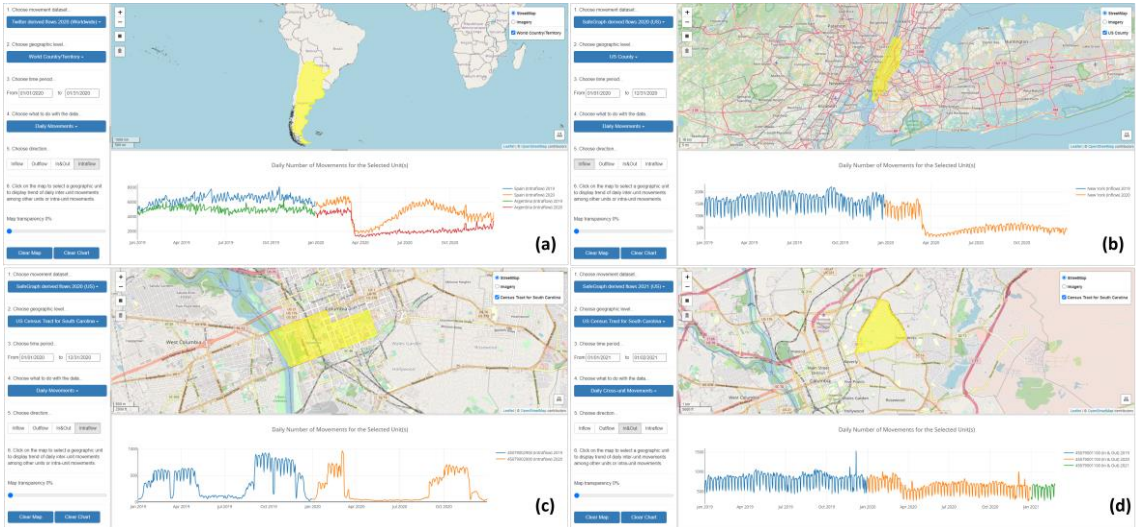

Figure 7. Daily population movement in different geographic scales. (a). Intraflow for Spain (top line) and Argentina (bottom line) in 2019 and 2020; (b) Inflow for New York County, U.S. in 2019 and 2020; (c) Intraflow for a census tract in Columbia, South Carolina (mainly located within the University of South Carolina) from 01/01/2019 to 02/24/2021; (d) Interflow (In&Out) for a census tract in a residential area of Columbia from 01/01/2019 to 02/24/2021.

Finally, the mobility data can be downloaded using the *Download* function by selecting interested data source, geographic level, time period, and aggregation type as csv (comma-separated values) files for further analysis. Each row in the csv file contains origin place (*o_place*), destination place (*d_place*), date (year, month, day if choosing *daily*), number of users/devices moved from origin to destination (*cnt*), and mean center of all flow origins (*o_lat, o_lon*) and flow destinations (*d_lat, d_lon*). Figure 8 shows over 160000 global daily flows were downloaded and visualized in the third-party mapping library kepler.gl (www.kepler.gl) as flow map (Figure 8b) and point density map (origin locations, Figure 8c). Section 4.3.2 details how to programmatically extract flow data and visualize in kepler.gl.

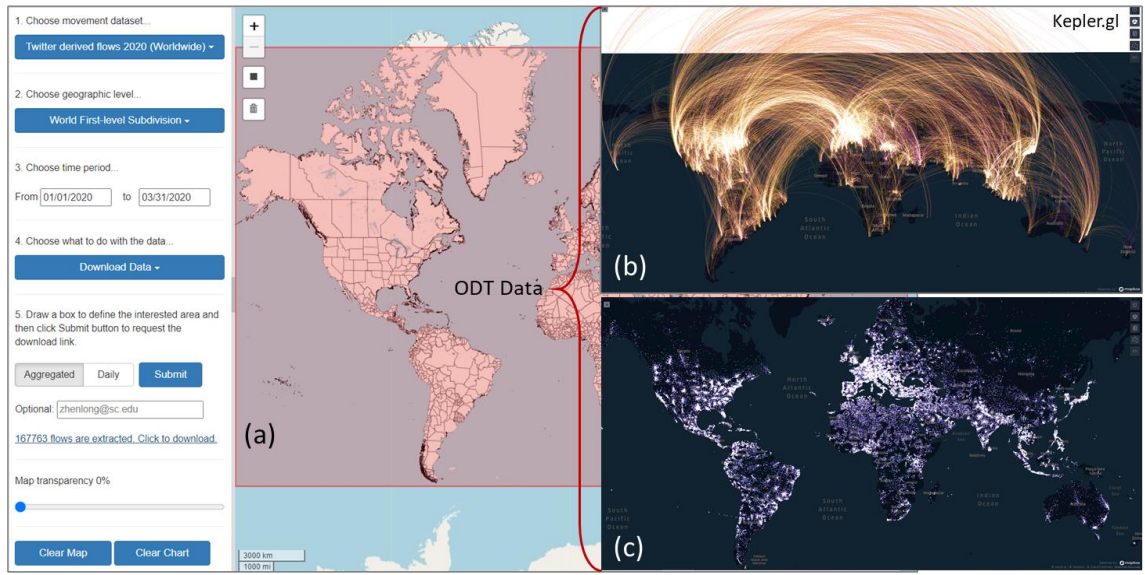

Figure 8. (a) Extract and download the Twitter-derived mobility data in ODT flow explorer at the world first-level subdivision from 01/01/2020 to 03/31/2020; (b) and (c): visualize the data in kepler.gl as flow map and point density map (using the origin locations).

**4.2 Integrating the ODT Flow APIs in the KNIME workflow environment**

To demonstrate how the ODT Flow APIs can be used with scientific workflow to produce reproducible, replicable, and expandable mobility data analysis, we developed two groups of case studies 1) dynamic map visualization using human mobility data; and 2) correlation analysis between human mobility and the COVID-19 infections. All case studies are developed using KNIME, a free and open-source visual workflow builder (KNIME, 2020).

**4.2.1 Dynamic map visualization using human mobility data**

The first case study used ODT Flow APIs to extract daily intra-state mobility values and visualize the mobility trends over time from March 1, 2020 to March 30, 2020 by dynamic choropleth maps. Figure 9 demonstrates visualization implementation by the scientific workflow tool KNIME. Each workflow consists of various types of nodes such as data reader, table joiner, python script, and line plot. In the demonstrated workflow, the procedures are presented intuitively, including 1) input data; 2) variables setting; 3) data processing; 4) map visualization; and 5) chart visualization. ODT APIs are used in the first step (Input Data) to programmatically extract the mobility data. More importantly, to make the data replicable, users only need to change input variables settings (e.g., change to different time periods) without extra operations. Figure 10 displays the interactive dynamic map visualization results and compares the intra-state human mobilities differences before and during the early stage of the COVID-19 pandemic in the U.S. The left map displays the choropleth map of intra-state human mobilities as of March 6, 2020, while the right map displays the mobilities choropleth map as of March 29, 2020. From both maps and line charts, it can be found human mobilities decreased in most of the states since mid-March 2020.

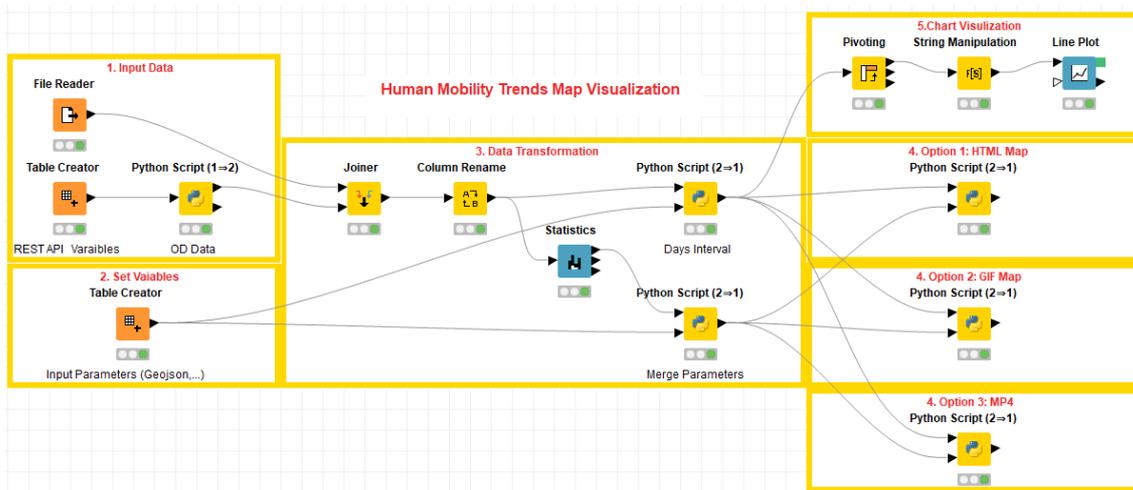

Figure 9. The workflow of human mobility intensity trends visualization.

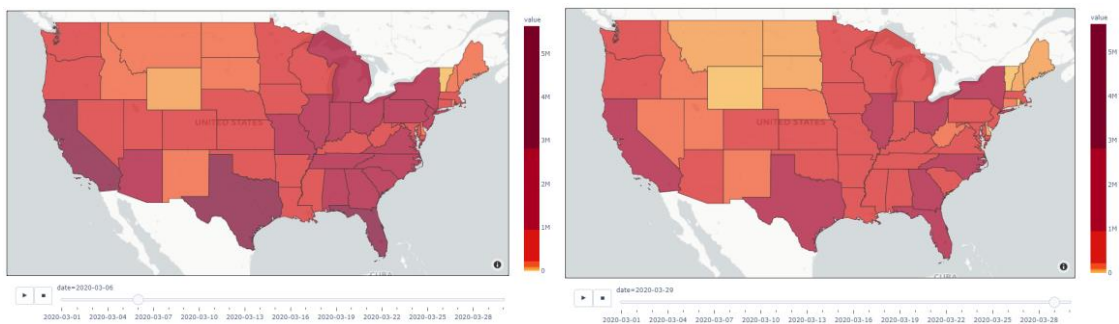

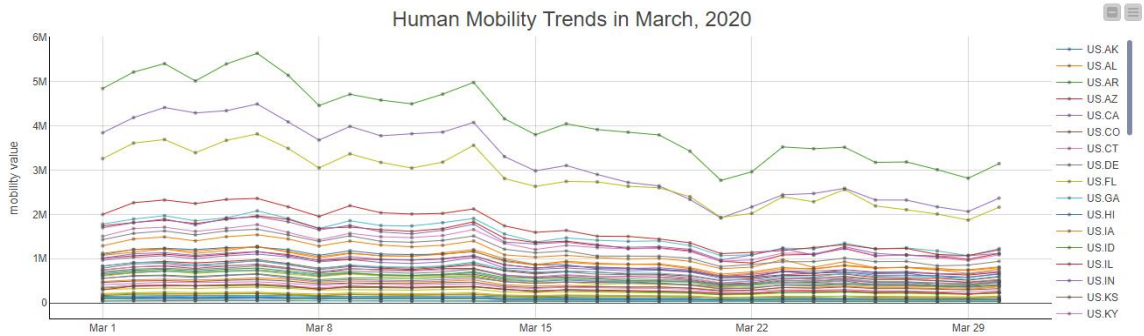

Figure 10. Visualization results of human mobility trends from the workflow.

### 4.2.2 Correlation analysis between human mobility and the COVID-19 infections

The magnitude and scale of human movement are critical for disease transmission prediction, risk area identification, and decision-making about control measures (Yang et al., 2020). The second case study builds a workflow, integrating population flows retrieved with ODT Flow APIs and COVID-19 infection cases data to investigate how human movements impact disease transmissions. New York state found its first case on March 1, 2020 and soon became the epicenter of the pandemic in the U.S., so we take New York as the study area to explore the impacts of New York population outflows on the COVID-19 spreading in other states. The U.S. confirmed cases are obtained from the shared COVID-19 datasets on Harvard Dataverse (Hu et al., 2020). Since New York was locked down on March 22, 2020, we compute the sum of outflows from New York to each state from March 20 through March 21, 2020 and estimate the daily correlation coefficients between total flows and confirmed cases. Furthermore, to evaluate the quality of Twitter-derived mobility data, we replicate the process using Safegraph-derived mobility data (only need to change the source variable of the API from *twitter* to *safegraph*) (Figure 11). The results from the workflow indicate that the correlation coefficients estimated by Twitter and Safegraph have the same trends that the values increase since March 22 and reach the peak (0.84 and 0.81) on March 30, which is 9 days to the lockdown day (Figure 12). The high correlations explain the close relations between human movement and COVID-19 virus transmissions. Besides, this case study also verifies the similarity of human mobility data derived from Twitter and Safegraph.

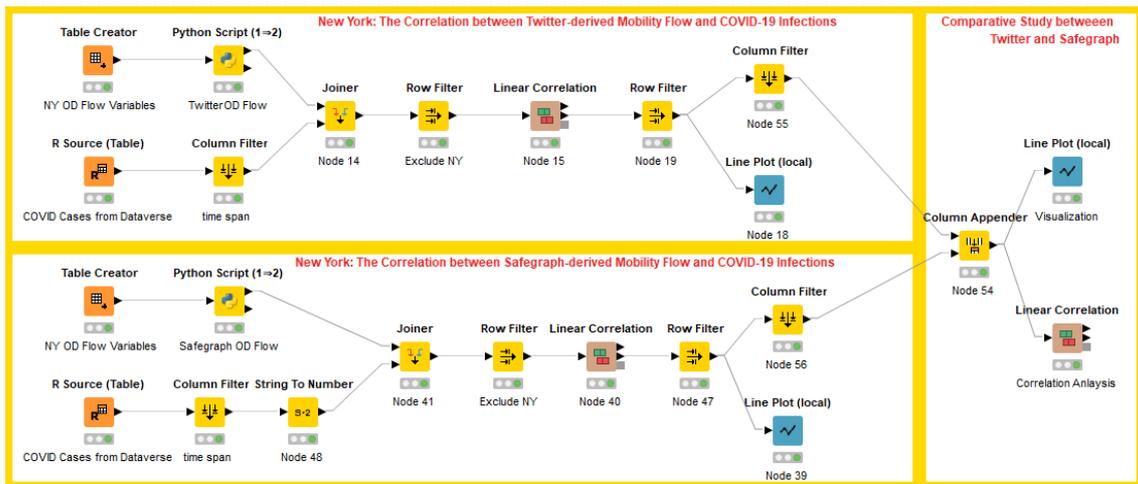

Figure 11. The correlation analysis of human mobility and COVID-19 infection cases in the U.S.

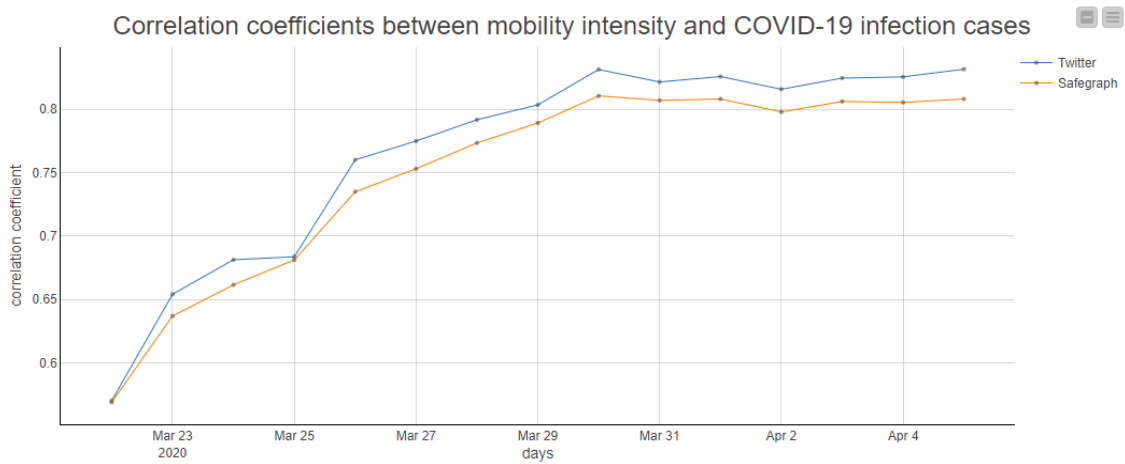

Figure 12. The correlation coefficients between outflows from New York and COVID-19 infection cases.

**4.3 Integrating the ODT Flow APIs in the Jupyter Notebook environment**

Jupyter Notebook (jupyter.org) is an open-source web application for interactive computing. It allows users to create and share documents that contain live code, running results, and comments. We show two case studies using Jupyter Notebook to access human mobility via ODT Flow APIs, and then conduct further analysis and visualization. In both case studies, the analysis from data fetching to results visualization can be completed in minutes using dozens of lines of Python code.

**4.3.1 Visual analytics of the impact of COVID-19 on human mobility in France**

In this case study, we visualize and analyze the impact of COVID -19 on human mobility in France's 13 administrative regions. Such spatial analyses could benefit policymaking at different jurisdictional levels. The first step is data preparation. Daily intra-flows of these 13 regions in 2020 were first extracted using the ODT Flow API. Figure 13 shows a sample code in this step for reading the boundary file (for mapping) and obtaining human mobility data using the API. We then computed the monthly change rates of flows compared with January 2020 for each region. Finally, the change rates of each month were rendered on maps, so that the spatial differences between regions can be visually investigated.

```
Read the boundary file

subdivision_file = r'gadm01_simplified/gadm36_1.shp'
gdf = gpd.read_file(subdivision_file)

target_place = r'FRA'    # set France as the target place (ISO code)
gdf_country = gdf[gdf['GID_1'].str[:3] == target_place] # Extract the boundary of the target place
```

```
Obtain 2020 flow data using the ODT Flow API

q = r'http://gis.cas.sc.edu/GeoAnalytics/REST'   #Set query url and parameters for the ODT REST API
params = {"operation": "get_daily_movement_for_all_places",
          "scale": "world_first_level_admin",
          "source": "twitter",
          "begin": "01/01/2020",
          "end": "12/31/2020"
         }
r = requests.get(q, params=params) #Submit request
df = pd.read_csv(StringIO(r.text))

df = df[df['place'].str[:3] == target_place] # Extract flows of the target place
```

Figure 13. Sample codes of reading the boundary file (for mapping) and obtaining human mobility data using the ODT Flow API

We compute the mobility reduction rate for each month of each region ($R_i$) with Eq.1 using the number of January 2020 flows ($M_{Jan.}$) as the baseline.

$$R_i = \frac{M_i - M_{Jan.}}{M_{Jan.}} \quad \text{Equation 1}$$

The maps in Figure 14 show the monthly mobility reduction rates of the 13 French regions. A nationwide drop in the intra-region mobility started to appear in March following the nationwide mandatory home lockdown on March 16, 2020 when France had over 5,500 COVID-19 cases and 127 deaths (Cuthbertson 2020) (Figure 15). The reduction rate peaked in April 2020 when most regions show over 50% of mobility reduction and started to decrease in May following the end of lockdown on May 11, 2020. In July and August, the mobility in southern regions was recovered along with the low daily infection cases. However, in November and December, the mobility in most regions of France experienced another dramatic reduction following the second nationwide lockdown on October 28 due to the second wave of infection and the high number of daily new death (Ledsom 2020). To visualize and analyze the mobility changes for another country, we only need to set *target_place* in the code to the country to be analyzed, and re-run the notebook to reproduce the maps, charts, and tables within a minute.

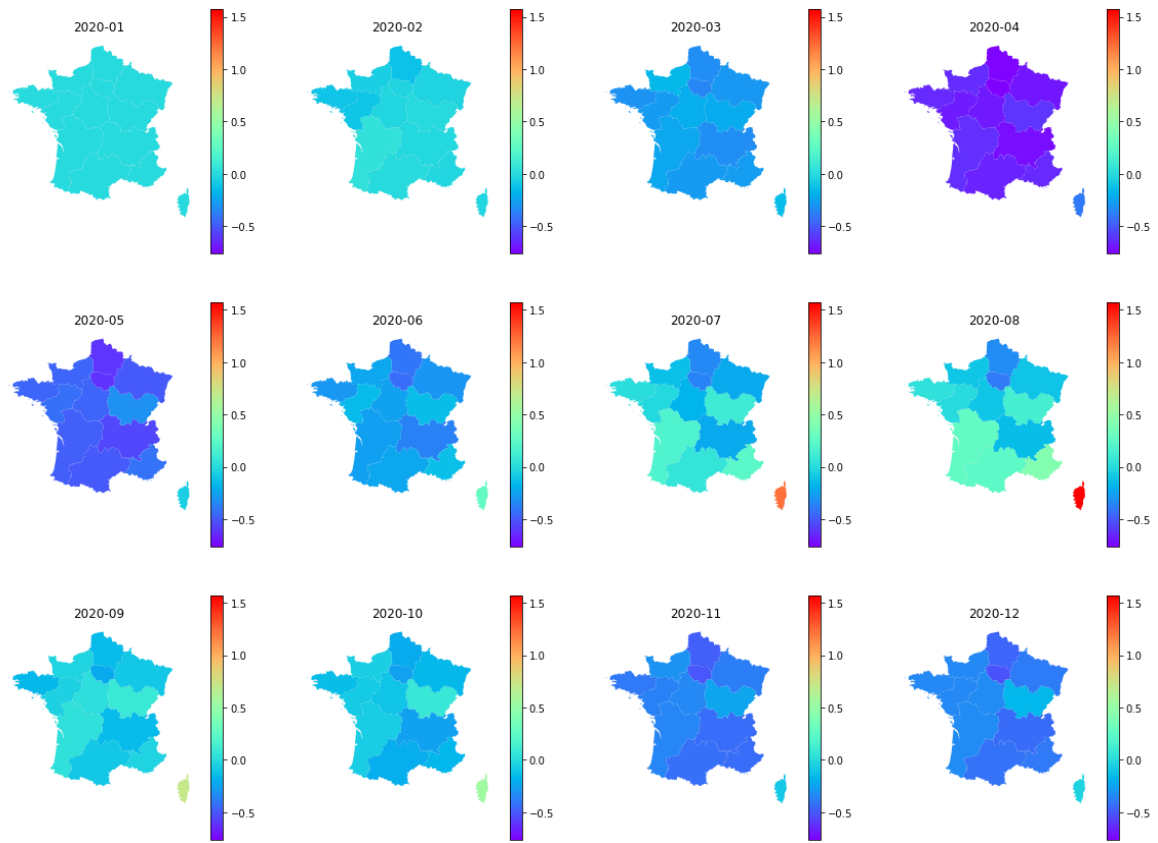

Figure 14 Monthly mobility change rates in 13 French administrative regions.

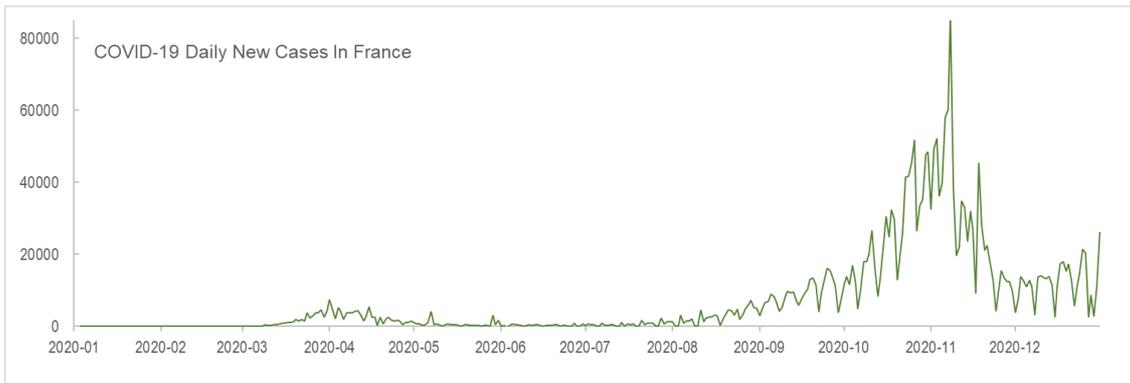

Figure 15 Daily COVID-19 new cases in France in 2020. (Data source: World Health Organization Coronavirus Dashboard, https://covid19.who.int/info/. Accessed on March 19, 2021)

**4.3.2 Interactive visualization of massive flows using ODT Flow APIs and kepler.gl**

The ODT Flow APIs allow users to on-demand query and extract large amounts of ODT flows at different geographic scales with designated study area (bounding box) and time period. As the extracted flow data contains latitude and longitude of the origin and destination locations, it can be directly loaded and visualized in third-party mapping libraries. Kepler.gl is an open-source WebGL-enabled high-performance mapping library, which supports interactive and responsive visualization of large location data. Figure 16 shows the entire code (except module import) of extracting the world first-level administrative OD flow matrix using the API and visualized as an interactive flow map using kepler.gl. The OD matrix is aggregated from January 1 to 5 in 2020 with the geographic area set to the whole world. New flow maps can be quickly regenerated by changing API parameters. For example, by changing the year from 2020 to 2019 and the type from "aggregated" to "daily", we can quickly generate a new world flow map showing 2019 daily flows. The daily flows can be further animated using kepler.gl's *Time Playback* function.

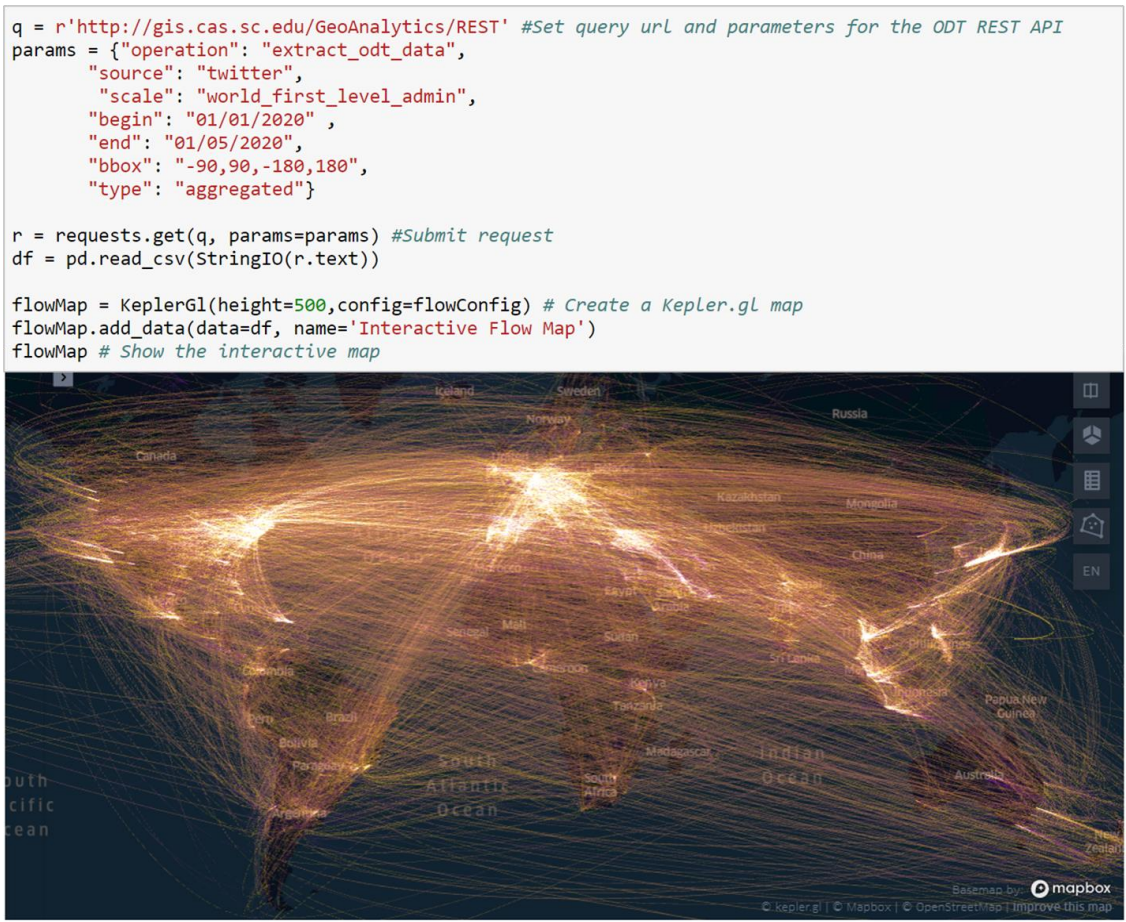

Figure 16. Python code of extracting the world first-level administrative OD flow matrix using the ODT Flow API and visualizing the data as an interactive flow map with kepler.gl

## 5. Discussions

### 5.1 How does the ODT Flow platform address the challenges of Big Data?

The overwhelming volumes in mobility datasets have been noted by many. In our case, the Twitter-derived and SafeGraph-derived daily OD Flows own 637 million and 11 billion Entity-O-D-T records, respectively. Such an amount demands not only massive data storage but also efficient processing techniques (intertwining with the *velocity* challenge). In the designed system, we aim to address the *volume/velocity* challenge by integrating our ODT data model with high-performance computing techniques. The datasets are stored and handled in a scalable big data computing environment, to facilitate efficient data processing, querying, extracting, and summarizing a large number of OD flows at the server-side. While the system is implemented on our in-house computing cluster, such a computing environment can be quickly provisioned in a cloud environment (e.g., Amazon EC2).

We tackle the *variety* challenge of the big mobility data via ODT cube, a unified conceptual data model, which supports the extraction and visualization of spatiotemporal standardized OD flows from heterogeneous movement data sources. In addition, the unified ODT cube allows multi-spatial scale queries (see Table 1) within a user-defined temporal period. The above features of the proposed ODT data model well respond to the multi-source and multi-scale characteristics of mobility data, originated from the *variety* challenge. In a similar manner, we aim to tackle the *veracity* challenge using the ODT model's capability to integrate multiple data sources in one unified framework, allowing for easy comparison and synthesis of mobility data at different spatiotemporal scales and capturing the multi-facets of human mobility, thus leading to a more comprehensive understanding of human movement that potentially benefits a wide range of research domains. While two types of mobility data sources (i.e., Twitter and SafeGraph, thanks to their cost-free availability) are included in this study, other mobility data sources can be further included and handled by the proposed framework in a similar way. Although different data sources require different human mobility extraction algorithms, other data sources, serving as plug-and-play components for the system, can be incorporated and implemented.

Numerous studies have demonstrated the important role (*value*) of mobility datasets in fields such as migration, urban planning, disaster management, to name a few. In this study, we direct our attention to demonstrating the utility of the proposed ODT framework in addressing the challenges in the ongoing COVID-19 pandemic (as of the time of writing). Our demonstrations reveal the similarity and dissimilarity of selected mobility sources and confirm close relations between human movement and COVID-19 virus transmissions. In this context, the proposed ODT Flow platform uniquely serves the soaring needs of human mobility data during disaster events such as the COVID-19 pandemic we are facing. The interactive web portal and APIs (integrated in KNIME workflow and Python Jupyter Notebook) meet the needs of different user communities, which maximizes the *value* of mobility data by reaching out to broader users.

**5.2 Reproducibility, replicability, and privacy**

There is a growing awareness of challenges in reproducibility and replicability facing the academic community (Choi et al., 2021; Kedron et al., 2021). The extensive use of locational data for applications such as disaster and humanitarian response raises the issue of reproducibility and replicability from competing perspectives of location privacy and geospatial data quality (Tullis and Kar, 2020). Following guidelines proposed by Choi et al. (2021), we aim to facility reproducibility and replicability by 1) providing a unified and well-documented online human mobility data repository and an interactive web portal that allows querying, visualizing, and downloading multi-source mobility data at different geographic scales, 2) providing APIs that facilitate easy programmatical access of the datasets to ease for reproducibility of studies the demand mobility data, and 3) providing mobility analysis demonstrations using the accessible and open computing environments of KNIME workflow and Jupyter Notebook.

Mobility data consists of individuals' location stamps, thus posing challenges to privacy protection as human movements are regarded as unique and predictable (De Montjoye et al., 2013). Scholars have voiced concerns on whether mobility data sharing is appropriate, even in the time of crisis like the COVID-19 pandemic (Bengio et al., 2020; Cho et al., 2020; Huang et al., 2020a). Nonetheless, mobility records that have been properly aggregated and anonymized (i.e., Google Community Mobility Reports, Apple Mobility Trends Reports, Descartes Mobility Records, and SafeGraph mobility records) have become acceptable and popular among academic communities (Huang et al., 2020b; Kogan et al., 2021; Change et al., 2021; Huang et al., 2021b; Gao et al., 2020). SafeGraph mobility records, one of the demonstrated mobility data sources in this study, are derived using a panel of GPS points from anonymous mobile devices. To enhance privacy, SafeGraph excludes cbg information if fewer than five devices visited an establishment in a month from a given cbg (SafeGraph, 2020). In our designed ODT framework, we further aggregate SafeGraph's records at the cbg-level to upper geographic scales, i.e., census tract, before sharing them with the public. As for Twitter data, we anonymize collected tweets and aggregate them to a geographic level as fine as the U.S. County. In future development, we will continue to follow standard privacy protection guidelines, ensuring mobility records regarding individuals' movements are protected.

**5.3 Data limitations**

In this study, the proposed ODT framework includes two data sources: 1) worldwide geotagged tweets collected using the Twitter public API (Twitter, 2021) and 2) Social Distancing Metrics (SDM) provided by SafeGraph based on U.S. mobile devices (SafeGraph, 2020). We acknowledge their intrinsic data limitations as below.

*Twitter-derived population flows*: The limitations of Twitter data have been documented by a number of studies (e.g., Li et al., 2013; Malik et al., 2015; Jiang et al., 2019). First, Twitter is not proportionally used by different population groups, thus presenting notable demographic and socioeconomic biases. For example, according to Statista (Tankovska, 2021), 28.4% of global Twitter users were aged between 35 and 49 years, 59.6% of global Twitter users were younger than 35, and 12% aged over 50. Second, geotagged tweets collected from the free public Twitter API (about 1% of the entire Twitter stream) are sparse and may not enough to capture the temporal patterns at the daily level for less populated areas. This is particularly the case when deriving county level daily population flows, as a Twitter user was included only when that user posted at least two tweets on a single day or posted tweets on at least two consecutive days. Third, the dynamics of people's Twitting activities (e.g., people tend to tweet more during big events), as well as the changing of Twitter's internal API, affect the daily number of tweets being collected. Thus, we advise that studies using the Twitter-derived flow data should be aware of these limitations when interpreting results and reaching conclusions. Nevertheless, a recent study suggests that Twitter-derived mobility data are able to capture the general human movement dynamics during the COVID-19 pandemic and present considerable similarity with other mobility data sources (Huang et al., 2021a).

*SafeGraph-derived population flows:* SafeGraph data have a high penetration rate (~10% of mobile devices in the U.S.) and well represent the U.S. population groups according to SafeGraph (2019). As a result, flows derived from SafeGraph are considerably denser than Twitter-derived flows, which overcomes the Twitter data limitations. Comparing to Twitter data, one downside of the SafeGraph-derived mobility data is their spatiotemporal scale, as SafeGraph data only date back to 2019, covering only the U.S. By enabling intuitive exploration and comparison of these two mobility datasets in the ODT Flow platform, this work highlights the importance and necessity of sharing and fusing multiple data sources for human mobility studies.

## 6. Conclusion

Human mobility dynamics provide fundamental knowledge regarding spatial interactions, benefiting a wide range of applications in need of such prior knowledge. The COVID-19 pandemic, to some degree, re-emphasizes the importance of human mobility monitoring and the value of mobility records. With the entering of the Big Data Era, new challenges start to appear, as fine-grained, large-scale human mobility records well fall into the category of Big Data, characterized by challenges of 5$V$s that demand a shift of data handling and sharing paradigm.

In response to the soaring needs of human mobility data, especially during disaster events such as the COVID-19 pandemic, and to the 5$V$ challenges in big mobility data, we develop a scalable platform for extracting, querying, visualizing, and sharing multi-source multi-scale human mobility data. The ODT data model is designed to work with parallel query engines (Apache Hive and Impala) to handle mobility data in large volumes with extensive spatial coverage. We process the human mobility data using a high-performance computing environment with source-specific human mobility extraction algorithms, which achieves efficient extracting of billion-level OD flows at the server-side. To enhance end-users' experience, we develop ODT Flow Explorer, allowing users to intuitively and interactively explore multi-source mobility datasets with user-defined spatiotemporal scales, which is expected to benefit both scientific communities and the general public in understanding human mobility dynamics. To promote reproducibility and replicability, we further develop ODT Flow REST APIs that provide researchers with the flexibility to access the data programmatically via workflows, codes, and programs. In the presented case studies, we demonstrate the potential of ODT Flow APIs coupled with KNIME scientific workflows and with Jupyter Notebooks to facilitate researchers to access and analyze massive OD flows.

While the ODT Flow platform (at the time of writing) features two mobility data sources, Twitter and SafeGraph, other mobility data sources, serving as plug-and-play components for the system, can be further included and handled by the proposed framework in a similar way. In future development, we will add more interactive visual analytics functions to the ODT Flow Explorer and consider replacing our in-house computing cluster to cloud computing environments, such as Amazon EC2.

**Data availability statement:** The derived multi-scale mobility data can be accessed via the interactive ODT Flow Explorer (http://gis.cas.sc.edu/GeoAnalytics/od.html) and the ODT Flow REST APIs. The following data, code, and materials are available on the project's GitHub page (https://github.com/GIBDUSC/ODT_Flow): 1) the scripts/program for constructing ODT flows from Twitter data and SafeGraph data using Apache Hive and Impala, 2) documentation and tutorial of the ODT Flow REST APIs, and 3) all source code of the case studies including the Jupyter Notebooks and KNIME workflows. A copy of the code is also available at Harvard Dataverse (https://doi.org/10.7910/DVN/GL3HAB). A video tutorial of the ODT Flow Explorer is available at https://www.youtube.com/watch?v=lV3AJIVYnSI.  Geotagged tweets were collected using Twitter's public Streaming API from the public domain following Twitter's Developer Agreement. Following Twitter's policy on "Redistribution of Twitter content" (https://developer.twitter.com/en/developer-terms/more-on-restricted-use-cases), the geotagged tweet IDs used in this analysis can be shared upon reasonable request. SafeGraph Social Distancing Metrics data are obtained from SafeGraph, Inc. (https://docs.safegraph.com/docs/social-distancing-metrics ).

**Acknowledgment:** The study and system development were supported by National Science Foundation (NSF) under grants 2028791 and 1937908, the National Institute of Allergy and Infectious Diseases (NIAID) of the National Institutes of Health (NIH) under grant R01AI127203-4S1, the University of South Carolina COVID-19 Internal Funding Initiative under grant 135400-20-54176, Leir Research Institute Faculty Seed Grant from New Jersey Institute of Technology, and start-up fund from Texas A&M University. The funders had no role in study design, data collection and analysis, or preparation of this article/system.

Kedron, P., Li, W., Fotheringham, S., & Goodchild, M. (2021). Reproducibility and replicability: opportunities and challenges for geospatial research. International Journal of Geographical Information Science, 35(3), 427-445.

Khan, N., Yaqoob, I., Hashem, I. A. T., Inayat, Z., Mahmoud Ali, W. K., Alam, M., ... & Gani, A. (2014). Big data: survey, technologies, opportunities, and challenges. The scientific world journal, 2014.

Kitamura, R., Chen, C., Pendyala, R. M., & Narayanan, R. (2000). Micro-simulation of daily activity-travel patterns for travel demand forecasting. Transportation, 27(1), 25-51.

KNIME. (2020). Accessed March 14, 20201. https://www.knime.com/

Kogan, N. E., Clemente, L., Liautaud, P., Kaashoek, J., Link, N. B., Nguyen, A. T., ... & Santillana, M. (2021). An early warning approach to monitor COVID-19 activity with multiple digital traces in near real time. Science advances, 7(10), eabd6989.

Kraak, M. J. (2003, August). The space-time cube revisited from a geovisualization perspective. In Proc. 21st International Cartographic Conference (pp. 1988-1996). Citeseer.

Kraemer, M. U., Yang, C. H., Gutierrez, B., Wu, C. H., Klein, B., Pigott, D. M., ... & Brownstein, J. S. (2020). The effect of human mobility and control measures on the COVID-19 epidemic in China. Science, 368(6490), 493-497.

Kumar, A., Gupta, P. K., & Srivastava, A. (2020). A review of modern technologies for tackling COVID-19 pandemic. Diabetes & Metabolic Syndrome: Clinical Research & Reviews, 14(4), 569-573.43

**Supporting Materials**

The derived multi-source multi-scale human mobility data are open sourced and can be accessed via the interactive ODT Flow Explorer at http://gis.cas.sc.edu/GeoAnalytics/od.html. A video tutorial showing how to use the ODT Flow Explorer is available at https://www.youtube.com/watch?v=lV3AJIVYnSI.

The open-sourced human mobility data can also be accessed programmatically via the ODT Flow REST APIs. The documentation of ODT Flow REST APIs is available at https://github.com/GIBDUSC/ODT_Flow/blob/main/ODT%20Flow%20REST%20APIs_Doc_v0.8.pdf

Tutorial (code examples) of how to access the ODT flow data programmatically using the ODT Flow REST APIs with Jupyter Notebook is available at https://github.com/GIBDUSC/ODT_Flows/blob/main/ODT%20Flow%20REST%20APIs_Notebook_Tutorial.ipynb

The source code (Jupyter Notebook) of the case study *Visual analytics of the impact of COVID-19 on human mobility in France* is available at https://github.com/GIBDUSC/ODT_Flow/tree/main/API%20with%20Jupyter%20Notebook%20Case%20study%201

The source code (Jupyter Notebook) of the case study *Interactive visualization of massive flows using ODT Flow APIs and kepler.gl* is available at https://github.com/GIBDUSC/ODT_Flow/tree/main/API%20with%20Jupyter%20Notebook%20Case%20study%202

The source code of the case studies with KNIME workflow (*Dynamic map visualization using human mobility data,* and *Correlation analysis between human mobility and the COVID-19 infections*) is available at https://github.com/GIBDUSC/ODT_Flow/tree/main/KNIME%20workflow%20case%20studies

The scripts/program for constructing ODT flows from Twitter data and SafeGraph data using Apache Hive and Impala is available at https://github.com/GIBDUSC/ODT_Flow

A copy of all the source code of the case studies including the Jupyter Notebooks and KNIME workflows is also available at Harvard Dataverse (https://doi.org/10.7910/DVN/GL3HAB).